\documentclass[osajnl,twocolumn,showpacs,superscriptaddress,10pt]{revtex4-1} 
\usepackage{amsmath,amssymb,graphicx}
\usepackage{color}
\usepackage{subfigure}
\begin{document}

\title{Detection of two-mode spatial quantum states of light by electro-optic integrated directional couplers}

\author{David Barral}\email{Corresponding author. Email: david.barral@usc.es}
\affiliation{Optics Area, Department of Applied Physics, Faculty of Physics and Faculty of Optics and Optometry, University of Santiago de Compostela, Campus Vida s/n (Campus Universitario Sur), E-15782 Santiago de Compostela, Galicia, Spain.}
\author{Mark G. Thompson}
\affiliation{Centre for Quantum Photonics, H. H. Wills Physics Laboratory \& Department of Electrical and Electronic Engineering, University of Bristol, Merchant Venturers Building, Bristol, UK.}
\author{Jes\'us Li$\tilde{\rm{n}}$ares}
\affiliation{Optics Area, Department of Applied Physics, Faculty of Physics and Faculty of Optics and Optometry, University of Santiago de Compostela, Campus Vida s/n (Campus Universitario Sur), E-15782 Santiago de Compostela, Galicia, Spain.}


\begin{abstract} We study both manipulation and detection of two-mode spatial quantum states of light by means of  a reconfigurable integrated device built in an electro-optical material in a Kolgelnik-Schmidt configuration, which provides higher error tolerance to fabrication defects and larger integration density than other current schemes.  SU(2) transformations are implemented on guided spatial modes in such a way that reconstruction of both the optical field-strength quantum probability distribution, via spatial two-mode homodyne detection, and the full optical field-strength wavefunction, by means of weak values, are carried out. This approach can easily be extended to spatial N-mode input quantum states. Apart from its usefulness to characterize optical quantum states, it is also emphasized its application to the measurement of the so-called generalized quantum polarization. 
\end{abstract}


\maketitle 

\section{Introduction}

\quad In recent years, great attention has been paid on the capabilities of light spatial modes in quantum mechanics. The technology based on them, integrated photonics, has lead to multiple new approaches on quantum communication, computation and sensing with huge success. These works are based on the processing of encoded quantum states of light in integrated waveguides, like single photons in a discrete space or squeezed light in a continuous-variable space, and measured by single-photon detectors or homodyne detection schemes, respectively \cite{OBrien2009}. 

The two main features for which integrated optics circuits are interesting to quantum information processing (QIP) are the sub-wavelength stability, which enables high-visibility quantum interference, and the great miniaturization they show with respect to bulk optics analogs, providing scalability \cite{Politi2008}. Likewise, the optical properties of the materials which make up the waveguides, are used for generating quantum states on chip, via their nonlinear features, manipulating them by means of their thermo-optic and electro-optic properties, among others, and even detecting these quantum states. 

As QIP technologies grow, quantum circuits become more complex and high-fidelity active control turns into an essential feature. Since the first demonstrations of quantum interference control based on the thermo-optic effect in optical waveguides \cite{Mathews2009,Smith2009}, this approach has been extensively adopted in Silicon-based ($Si$) quantum circuits with notable success \cite{Shadbolt2012,Bonneau2012a,Metcalf2014,Silverstone2014}. Besides, quantum interference has also been shown recently by means of strain-optic based phase controllers in Silica ($SiO_{2}$) \cite{Humphreys2014}. The other main branch, electro-optic based quantum circuits, exhibits very promising attributes as well. Lithium Niobate ($LiNbO_{3}$) photonic circuits have demonstrated efficient generation of entangled photons and fast control on chip \cite{Martin2012, Jin2014}, even on both polarization and path degrees of freeedom \cite{Bonneau2012}, and storage as a quantum memory \cite{Saglamyurek2011}. Additionally, Gallium Arsenide ($GaAs$) is another high-performance material capable to generate \cite{Santori2002}, manipulate \cite{Wang2014} and measure \cite{Sahin2015} single photons in photonic circuits. These features position electro-optic materials in a pre-eminent place for future quantum photonics technologies, where high integration and fast and precise modulation will be required.

Both the processing and measuring of quantum states, like those carried out in the works presented above, are based on unitary transformations, since they leave all physical predictions of quantum mechanics invariant. When we are dealing with qubits, they are accomplished as rotations in the Bloch sphere like in discete-variable quantum computation (DVQC) \cite{Nielsen2010}. Likewise, in the case of looking for full characterization of a quantum state, they appear as rotations in the phase space. This is known as quantum state tomography (QST) and is based on homodyne detection \cite{Lvovsky2008}. This technique enables us to extract all the information of the quantum state reconstructing the Wigner function and the density operator in the quadrature, Fock, computational spaces and so on \cite{Shadbolt2012, Smithey1993, Opatrny1997,Raymer1996}. Another way of quantum state reconstruction is based on weak values \cite{Aharonov1988, Lundeen2011}, where measurement is carried out by imposing postselection of the quantum state and weak interaction between the measurement apparatus and the state itself, leaving the state of interest largely undisturbed. In a similar way, properties of $N$-dimensional quantum states propagating in quantum circuits have been studied in the optical field-strength $\bold{\mathcal{E}}$. In this space, the quantum wavefunction presents a generalized polarization which can be quantifed by a degree of polarization and measured by homodyne detection techniques as well \cite{Linares2011, Barral2013, Luis2013}. Additionally, coherent detection is also carried out in continuous-variable quantum computation (CVQC) \cite{Braunstein2005}.

As manipulation and characterization of quantum states of light in integrated photonics is a very active field of research by all the aforementioned, we introduce a reconfigurable integrated device capable to carry out SU(2) and SO(2,R) operations based on the Kogelnik-Schmidt scheme for electro-optic directional couplers (DC) \cite{Kogelnik1976}. The advantages of this proposal are on one hand fast modulation, based on the electro-optic nature of the material, and on the other hand higher error tolerance to fabrication defects and larger integration density than other current schemes due to the design of the device. Its architecture enables carrying out both SU(2) transformations and fabrication defects correction simultaneously, as well as the number of elements is less than in other proposals. All this is a very important fact because as the complexity and number of elements of a photonic circuit network increase, the effect of imperfections in its operation becomes more problematic. Moreover, the integrated nature of our scheme gives access to bigger dimension quantum states by means of nesting. In addition, we propose two original applications: firstly, the measurement and characterization of spatial quantum states of light by means of optical-field strength homodyne detection and secondly, two-dimensional wavefunction reconstruction by means of weak measurements. However, this device can be also applied in QST, DVQC, CVQC or any other task where orthogonal or unitary operations are necessary.


Briefly, the paper is organized as follows. In Section 2 we present the formalism of propagation of quantum states of light in integrated devices based on the Momentum operator and introduce the optical field-strength representation. In Section 3 we sketch a reconfigurable directional coupler as a device which carries out SU(2) and SO(2,R) transformations in two-mode input quantum states and as a component of a SU(N) / SO(N,R) device, and we compare its performance with that of other current schemes. In Section 4 we focus on possible applications of this device. In the first place we use this device as part of an homodyne detector for optical-field strength measurements of quantum light, and next we apply it to weak measurements of a two-dimensional wavefunction. Finally a summary is presented.

\section{Propagation of quantum light in integrated waveguides}

It is well-known that the generator of spatial propagation in quantum mechanics is the dynamical Momentum operator $\hat{M}$ \cite{Luks2002}. In an integrated photonic device composed by $N$ coupled linear single-mode homogeneous waveguides it is given by \cite{Linares2008}:
\begin{equation}\label{Momentum}
\hat {M}=\sum_{\sigma=1}^{N} \hbar\,\tilde{\beta}_{\sigma}\,\hat{a}_{\sigma}^{\dag}\hat{a}_{\sigma}^{}+\sum_{\sigma<\sigma'}\{\hbar \,\kappa_{\sigma, \sigma'} \,\,\hat{a}_{\sigma}^{}\hat{a}_{\sigma'}^{\dag}+h.c.\},
\end{equation}
where $\tilde{\beta}_{\sigma}=\beta_{\sigma} + \kappa_{\sigma, \sigma}$,  $\beta_{\sigma} $ is the propagation constant of the $\sigma$-mode, where $\sigma$ stands for the modal numbers $\nu$, $\mu$ in each transverse direction, $\kappa_{\sigma,\sigma}$ is the self-coupling coefficient, $\kappa_{\sigma, \sigma'}$, where $\sigma\neq\sigma'$, the cross-coupling coefficient, and $h.c.$ stands for hermitian conjugate. 
$\hat{a}_{\sigma}\,(\hat{a}_{\sigma}^{\dag})$ is the usual spatial absorption (emission) operator fulfilling the equal space commutation relation,
\begin{equation} \label{Commutation}
[\hat{a}_{\sigma}(z),\hat{a}_{\sigma'}^{\dag}(z)]=\delta_{\sigma,\sigma'},
\end{equation}
and Heisenberg's equations which describe the propagation of quantum states of light, given by:
\begin{equation}\label{Heisenberg}
\frac{d \hat{a}_{\sigma}}{dz} = \frac{i}{\hbar} [\hat{a}_{\sigma}, \hat{M}_{\sigma}].
\end{equation}
These operators are central in quantum optics because their eigenstates are the coherent states $\vert \alpha_{\sigma} \rangle$. Moreover, from these operators we build the number operator $\hat{n}_{\sigma}=\hat{a}_{\sigma}^{\dag}\hat{a}_{\sigma}$, with the Fock states $\vert n_{\sigma} \rangle$ as eigenstates, or the optical field-strength operator $\hat{\mathcal{E}}_{\sigma}=(\hat{a}_{\sigma}+\hat{a}_{\sigma}^{\dag})/2$ fulfilling $\hat{\mathcal{E}}_{\sigma}\vert \mathcal{E}_{\sigma} \rangle={\mathcal{E}}_{\sigma}\vert \mathcal{E}_{\sigma} \rangle$, with $\vert \mathcal{E}_{\sigma} \rangle$ the optical field states, among many others. In particular these operators are very important because any quantum state can be expressed as linear combinations of their eigenstates. Thus, different problems or applications in quantum optics are better suited to different spaces, as the Fock basis for DVQC \cite{Politi2008} or the optical field-strength basis for the study of quantum polarization \cite{Linares2011}. In this work we will center our attention to this last space since it is the natural one for the problem of homodyne detection, though our approach could work in any of the above-mentioned representations.

Any multimode quantum state of light $\vert L\rangle$ can be written as a superposition of eigenstates $\vert \mathcal{E}_{1}, ... , \mathcal{E}_{N}\rangle$ of the optical field-strength operators $\hat{\mathcal{E}}_{\sigma}$, 
\begin{equation}\label{WFe}
\vert L\rangle=\int \langle \mathcal{E}_{1}, ... ,\mathcal{E}_{N}\vert L\rangle \vert \mathcal{E}_{1}, ... ,\mathcal{E}_{N}\rangle \, d \mathcal{E}_{1}... d \mathcal{E}_{N},
\end{equation}
with $ \langle \mathcal{E}_{1}, ... , \mathcal{E}_{N}\vert L\rangle\equiv\Psi(\mathcal{E}_{1}, ... , \mathcal{E}_{N})=\vert \Psi (\mathcal{E}_{1}, ... , \mathcal{E}_{N})\vert\,e^{i\,\varphi(\mathcal{E}_{1}, ... , \mathcal{E}_{N})}$ the complex probability amplitude or wavefunction in the optical field-strength domain. Hence the probability distribution on the $N$-dimensional optical field-strength space $\boldsymbol{\mathcal{E}}\equiv(\mathcal{E}_{1}, ... ,\mathcal{E}_{N})$, is given by:
\begin{equation}
P(\mathcal{E}_{1}, ... , \mathcal{E}_{N})= \vert  \langle \mathcal{E}_{1}, ... , \mathcal{E}_{N}\vert L\rangle \vert^{2}\equiv\vert \Psi (\mathcal{E}_{1} , ... , \mathcal{E}_{N})\vert^{2}.
\end{equation}
The shape of these probability distributions is directly related to the generalized polarization degree of the quantum state analyzed, that is, to the confinement properties of its probability distribution in the optical field-strength space \cite{Linares2011, Barral2013}. Alternatively, some problems are better suited to be tackled in the conjugate quadrature, the canonical optical momentum $\boldsymbol{\mathcal{P}}$, fulfilling:
\begin{equation}\label{P}
[\hat{\mathcal{E}}_{\sigma}(z),\hat{\mathcal{P}}_{\sigma'}(z)]=i\delta_{\sigma,\sigma'}/2.
\end{equation}
The Fourier transform relates the complex wave function in the optical field-strength and optical momentum domains. Besides, the propagation of these complex probability amplitudes are suitable to be worked out by means of an spatial propagator \cite{Linares2012}, which simplifies the calculation in some problems. 

\begin{figure}[h]
\centering
\includegraphics[width=0.48\textwidth]{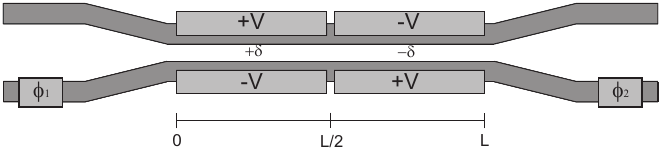}
\vspace {-0.3cm}\,
\hspace{-2cm}\caption{\label{F1}\small{Sketch of the device proposed.}}
\end {figure}

\section{Reconfigurable directional couplers}

As it was commented above, high integration density, fidelity and fast control are main goals in future quantum photonic devices. On one hand we have Silicon on insulator technology (SOI), which has exploited the high refractive index contrast and large thermo-optic coefficient of Silicon, demonstrating good performance in QIP \cite{Shadbolt2012,Bonneau2012a,Metcalf2014,Silverstone2014}. The main drawback of these devices is a typical length of several hundred micrometers, as they suffer from large power dissipation and therefore thermal crosstalk. Fast modulation can be obtained by means of integrated forward-biased pin-junctions or reverse-biased pn-junctions into the SOI waveguide, but suffer from high energy cost and low phase control efficiency, respectively \cite{Pfeifle2012}. On the other hand we have the electro-optic-based technology, as $LiNbO_{3}$ and $GaAs$ photonics. High bandwidths, precise control and integration density can be obtained with this technology. Electro-optic efficiency and bias voltage drift were some of the problems these materials showed in the past, but via engineered solutions they present nowadays an excellent behaviour \cite{Wooten2000, Walker2012}, and they are making their way in QIP \cite{Martin2012, Jin2014, Bonneau2012, Saglamyurek2011, Santori2002, Wang2014, Sahin2015}. 

The advent of this technology and the need of active control on quantum states for manipulation and measurement, leads us to propose a quantum photonic device which enables reconfigurable SU(2) and SO(2,R) transformations based on a directional coupler with two-section reversed electrodes, that is, an alternating $\Delta \beta$-coupler or Kogelnik-Schmidt scheme \cite{Kogelnik1976}, {together with} two electro-optic phase shifters. To our knowledge this is the first report of an electro-optic reconfigurable circuit which performs any unitary operation on spatial modes of quantum light, as previous works dealt only with control of the path photons take \cite{Martin2012, Jin2014}. In the case of polarization-encoded quantum light, a reconfigurable unitary device has been previously described \cite{Bonneau2012}. We present below the scheme of the device proposed.

\subsection{The device}

It is known that unitarity is a restriction on the allowed operations in quantum mechanics \cite{Nielsen2010}. Any discrete unitary transformation of a two-mode input state, that is an U(2) transformation, can be accomplished experimentally in bulk optics by means of a beam splitter with variable reflectivity and a phase shifter at one output port or, alternatively, substituting the beam splitter by a Mach-Zehnder interferometer \cite{Reck1994}. In integrated photonics this last approach has been recently shown using two phase shifters and two passive 3dB DCs \cite{Shadbolt2012}. Here we adopt the first scheme above introduced by means of an electro-optic DC.


\begin{figure}[h]
\centering
\includegraphics[width=0.5\textwidth]{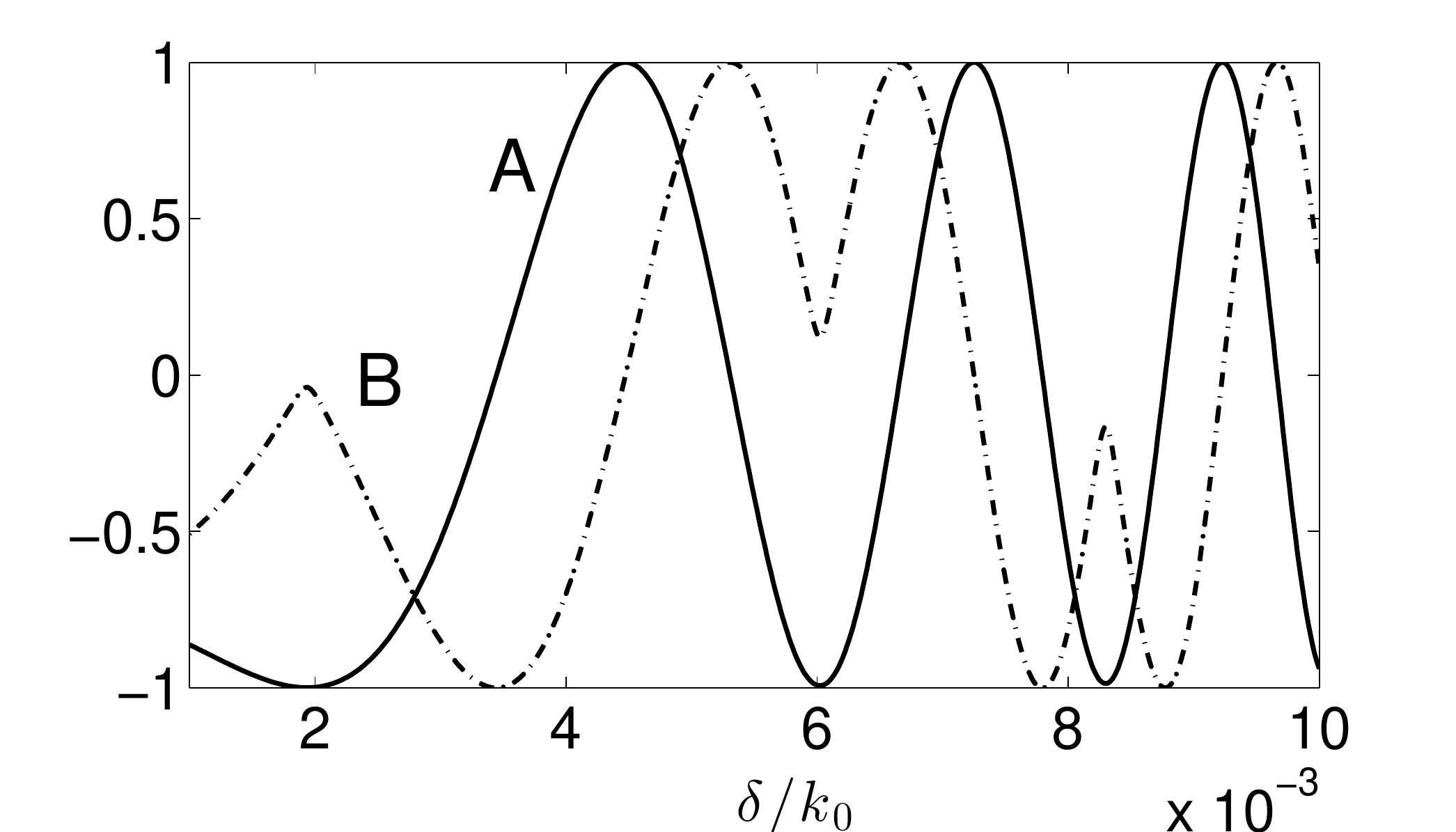}
\vspace {-0.2cm}\,
\hspace{-2cm}\caption{\label{F2}\small{    Values of $A$ (solid line) and $B$ (dash-dot line) versus $\delta/k_{0}$ for a directional coupler with parameters $\kappa=0.1 k_{0}$ m$^{-1}$, $L=2$ mm and wavelength $\lambda=650$ nm.    }}
\end {figure}

\begin{figure}[h]
  \centering
    \subfigure{\includegraphics[width=0.5\textwidth]{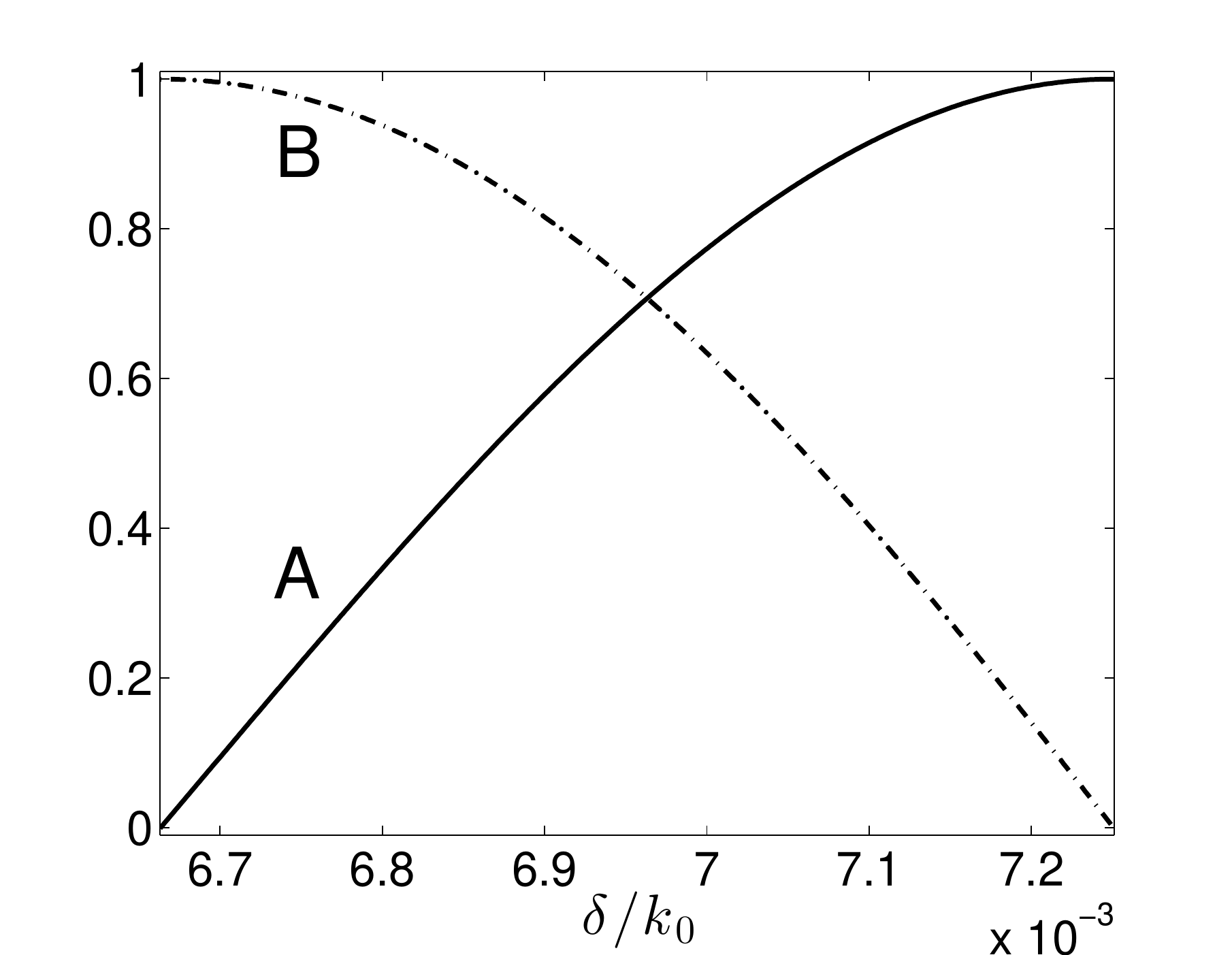}}
    \subfigure{\includegraphics[width=0.5\textwidth]{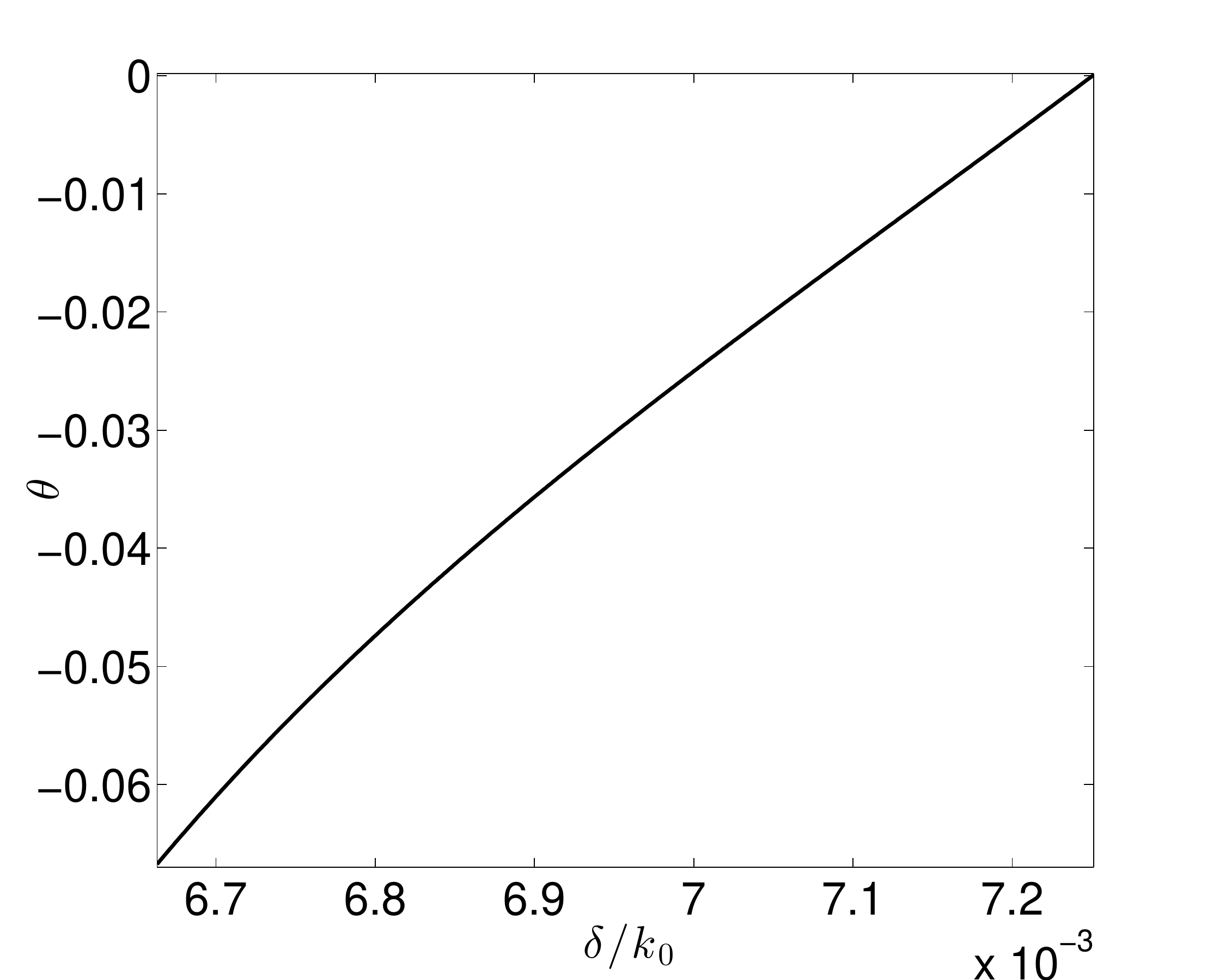}}
    \vspace {-0.3cm}\,
\hspace{-2cm}\caption{\label{F3}\small{Values of A (solid line, upper figure), B (dash-dot line, upper figure) and $\theta$ (lower figure) versus $\delta/k_{0}$ for a directional coupler with parameters $\kappa=0.1 k_{0}$ m$^{-1}$, $L=2$ mm and wavelength $\lambda=650$ nm.  }}
\end{figure}

\begin{figure*}[t]
\begin{center}
\includegraphics[width=0.85\textwidth]{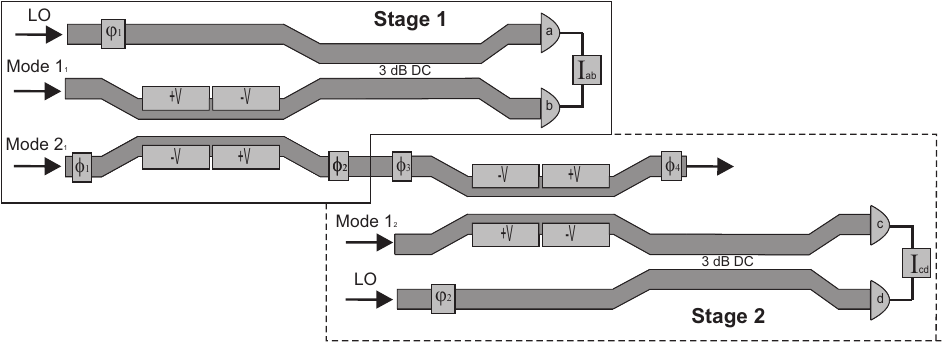}
\vspace {-0.3cm}\,
\hspace{-2cm}\caption{\label{F5}\small{Homodyne detection scheme for measurement of generalized polarization of a three-mode input quantum state of light composed by two measurement units, stages 1 and 2. Looking at stage 1 (solid line), $1_{1}$ and $2_{1}$ modes are sent to the Kogelnik-Schmidt coupler where rotations in the field-strength space $\boldsymbol{\mathcal{E}}$ are performed. The upper output mode is then mixed in a BHD with a LO-mode. The lower ouput mode acts as the upper input mode ($2_{2}$) of the Kogelnik-Schmidt coupler in the next stage (dashed line), where it is mixed with mode $1_{2}$. The free output of this coupler would act as input in the next stage and so on $N-1$ times for $N$ modes.}}
\end{center}
\end {figure*}

Our device is depicted in Figure 1. The initial and final stages of the device are electro-optic phase shifters, meanwhile the central part is made up of a DC of length $L$ with two sections of reversed electrodes over it \cite{Kogelnik1976}. The DC is composed of two asynchronous waveguides where TE polarized light propagates. Optical modes $1$ and $2$ propagate in the device with propagation constants $\beta_{1}$ and $\beta_{2}$, respectively. $L/2$ are the lengths of the two reversed electrodes, $\kappa$ is the coupling coefficient of the DC, $\Delta \beta (V)=(\beta_{1}-\beta_{2})\equiv 2\delta$ is the propagation constant mismatch between the waveguides to be modulated electro-optically by the voltage $V$ \cite{Yariv1975} and $\phi_{1}(V')$, $\phi_{2}(V'')$ are input and output electro-optic phase shifts controlled by two additional electrodes with voltages $V'$ and $V''$, respectively. Solving Heisenberg's equations (\ref{Heisenberg}) after applying the Momentum operator which describes this system (\ref{Momentum}), we obtain the following transformation performed by the device:
\begin{equation} \label{M}
M(\delta, \phi_{1}, \phi_{2})=
  \begin{pmatrix}
     A(\delta) &  iB(\delta)\,e^{i(\theta+\phi_{2})} \\
     iB(\delta)\,e^{-i(\theta-\phi_{1})} &  A(\delta)\,e^{i(\phi_{1}+\phi_{2})}
  \end{pmatrix},
\end{equation}
where $A=u^{2}-v^{2}$, $B=2uv$, and the functions $u$, $v$ and $\theta$:
\begin{eqnarray} \label{uv}
u=[{\cos}^{2}\beta_{r}L + \frac{\delta^{2}}{\beta_{r}^{2}}{\sin}^{2}\beta_{r}L]^{1/2}, \,\, v=\frac{\vert \kappa \vert}{\beta_{r}} {\sin}\beta_{r}L, \\ \label{thetas}
\theta={\rm atan}(\frac{\delta}{\beta_{r}}\,{\tan}\beta_{r}L),
\end{eqnarray}
with $\beta_{r}=[\kappa^{2} + \delta^{2}]^{1/2}$ and where $u^{2}+v^{2}=1$. In Figure 2 we plot the functions A and B for different values of $\delta/{k_{0}}$, with $k_{0}$ the propagation constant in vacuum for a given wavelength $\lambda$. These functions differ from the cosine and sine at some values, for instance at $\delta/{k_{0}}\approx 6.\,10^{-3}$ in Figure 2, and their rotation speed increases with $\delta$. To perform the rotation we only need angles between $0$ and $\pi/2$, so we can thoroughfully choose a range of values of $\delta/{k_{0}}$ where A and B mimic the cosine and sine functions, respectively. The change in the velocity of rotation does not represent a problem in the experimental regime as the values of $\delta$ are chosen from discrete equally separated values of A and B preselected by the user.
The phase shifters $\phi_{1}$ and $\phi_{2}$ can be adjusted in different ways to accomplish the desired operation. Choosing the next simple way:
\begin{equation} \label{Phis}
\phi_{1}=\Phi+\theta+\pi/2=-\phi_{2}, 
\end{equation}
we obtain the following transformation in equation (\ref{M}):
\begin{equation} \label{M2}
M(\delta, \Phi)=
  \begin{pmatrix}
       A(\delta)&  B(\delta) e^{-i\Phi} \\
     - B(\delta) e^{i\Phi} & A(\delta)
  \end{pmatrix}.
\end{equation}
For the sake of clarity, note that if we take $u= \cos(\Theta(\delta)/4)$ and $v= \sin(\Theta(\delta)/4)$,  we can rewrite A and B as:
\begin{equation}
A=\cos(\Theta(\delta)/2), \quad B=\sin(\Theta(\delta)/2).
\end{equation}
Hence, the transformation given by equation (\ref{M2}) is an effective SU(2) unitary up to a global phase without physical significance, and a rotation SO(2,R) if $\Phi=n\pi$ is chosen. Then, to get an arbitrary unitary transformation or equally, to select $\Theta$ and $\Phi$, firstly $\delta$ is set by means of the electrodes voltage $V$ to choose the desired $\Theta$. This assigns $\theta$ a value (\ref{thetas}) used in equations (\ref{Phis}) to adjust the electrode voltages of the phase shifters $\phi_{1,2}$, and obtain the sought value of $\Phi$. All these adjustments would be controlled continously by a computer. In Figure 3 we show the values of A, B (upper Figure) and $\theta$ (lower Figure), respectively, versus $\delta/{k_{0}}$ for a bandwith $\delta$ completing a rotation. 

On the other hand, any $N$-mode quantum state can be manipulated and measured by means of nesting an array of devices like that depicted in Figure 1. U(N) transformations can be carried out by means of succesive U(2) operations on two-dimensional subspaces leaving an $(N-2)$-dimensional subspace unchanged, with $N>2$ \cite{Reck1994}. We can carry out this control on a computational basis, quadrature basis and so on, thus becoming a useful device in quite different quantum optics areas. In the next Section we will apply it to optical quantum measurement, in particular via homodyne and weak values detection. 



\subsection{Performance of the device}

Fast modulation and fidelity are significant features of current electro-optic devices. For example, in the case of $LiNbO_{3}$ used as material support, switching bandwiths of $40$ GHz are available in telecommunications commercial modulators as well as $100$ GHz modulation has been achieved in the laboratory \cite{Kanno2010}; likewise, long-term field reliability has been demonstrated \cite{Wooten2000}. But the main improvement of this design over other current schemes is its ability to get over fabrication errors of the couplers, related for instance with the coupling constant $\kappa$ or the coupling length $L$, by simply adjusting the electrodes voltage \cite{Kogelnik1976}. To demonstrate this, we show what happens in the case of fabrication imperfections in the current integrated SU(2) scheme, that made up of a Mach-Zehnder interferometer (MZI), that is two $3 dB$ passive DC with an active phase shifter in between, followed by a second active phase shifter \cite{Shadbolt2012,Bonneau2012a,Metcalf2014,Silverstone2014,Jin2014}. The scattering matrix of a $3 dB$ DC with fabrication defects is given by:
\begin{equation} \label{ErrorDC}
\frac{1}{\sqrt{2}}
  \begin{pmatrix}
       1-\epsilon &  i(1+\epsilon) \\
     i(1+\epsilon) & 1-\epsilon
  \end{pmatrix},
\end{equation}
with $\epsilon$ a parameter standing for the defects and where only first order errors $O(\epsilon)$ have been taken into account. An integrated Mach-Zehnder made up of two $3dB$ DCs with defects $\epsilon$ and $\epsilon '$ as in equation (\ref{ErrorDC}), respectively, and a phase shifter causing a change $\eta$, is given by the next transformation:
\begin{equation} \label{ErrorMZ}
  \begin{pmatrix}
       \cos(\eta)+i(\epsilon+\epsilon')\sin(\eta) &  \sin(\eta)+i(\epsilon'-\epsilon)\cos(\eta) \\
     \sin(\eta)-i(\epsilon'-\epsilon)\cos(\eta) & -\cos(\eta)+i(\epsilon+\epsilon')\sin(\eta)
  \end{pmatrix},
\end{equation}
where a global phase has been dismissed. So the scattering matrix of the MZI becomes complex in the case of fabrication errors, which can not be compensated by the output port phase shifter and therefore accurate $SU(2)$ transformations can not be accomplished \cite{Metcalf2014}. In our case, this is easily solved by adjusting the values of $\delta(V)$, $\phi_{1}(V')$ and $\phi_{2}(V'')$, as it can be seen by inspection of equations (\ref{uv}-\ref{M2}). Moreover, the above statement could be refuted by using alternating  $\Delta \beta$ couplers in the MZI, but in that case our design saves one DC, improving the integration density. It is important to outline that in complex networks the larger the number of DCs are involved the higher the probability of fabrication imperfections we have, and therefore larger deviation from an  ideal operation. Hence the importance of this scheme in future QIP technologies.

\section{Applications to  optical  quantum detection}

In this section we show how our electro-optic SU(2) device can be used to measure  the optical field-strength probability distribution  of a two-mode quantum state by homodyne detection, that is, $\vert\Psi(\mathcal{E}_{1},\mathcal{E}_{2})\vert^{2}$,  by means of rotations in the optical field-strength space $\boldsymbol{\mathcal{E}}$. Likewise, we extend it to the full reconstruction of the wavefunction, that is getting amplitude and phase of a two-mode quantum state by using weak values detection, that is, $\Psi(\mathcal{E}_{1},\mathcal{E}_{2})$.  For sake of both simplicity and clarity we present the applications with two-mode quantum states although it can be easily extended to N-mode quantum states.

\begin{figure}[h]
  \centering
    \subfigure{\includegraphics[width=0.55\textwidth]{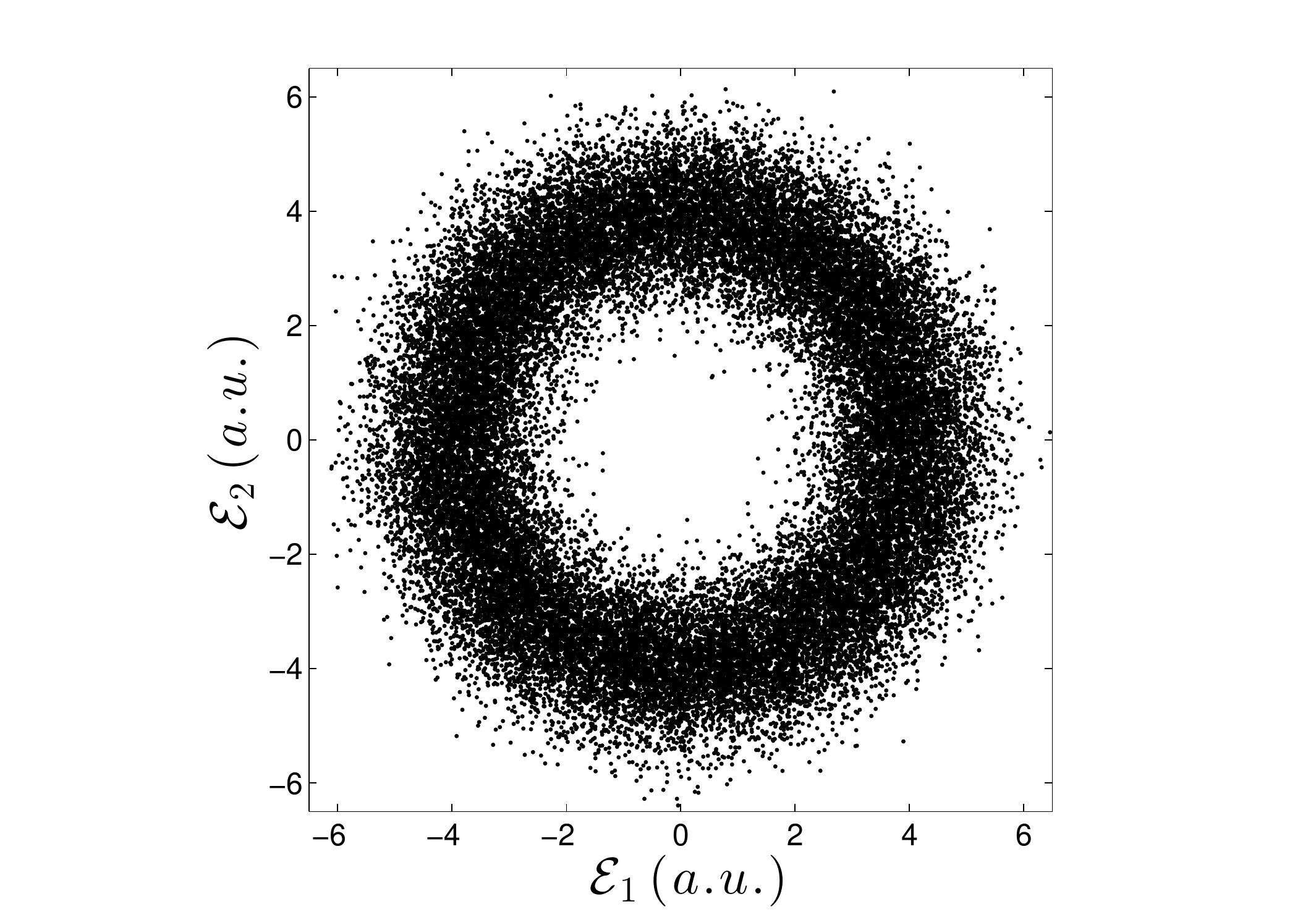}}
    \subfigure{\includegraphics[width=0.52\textwidth]{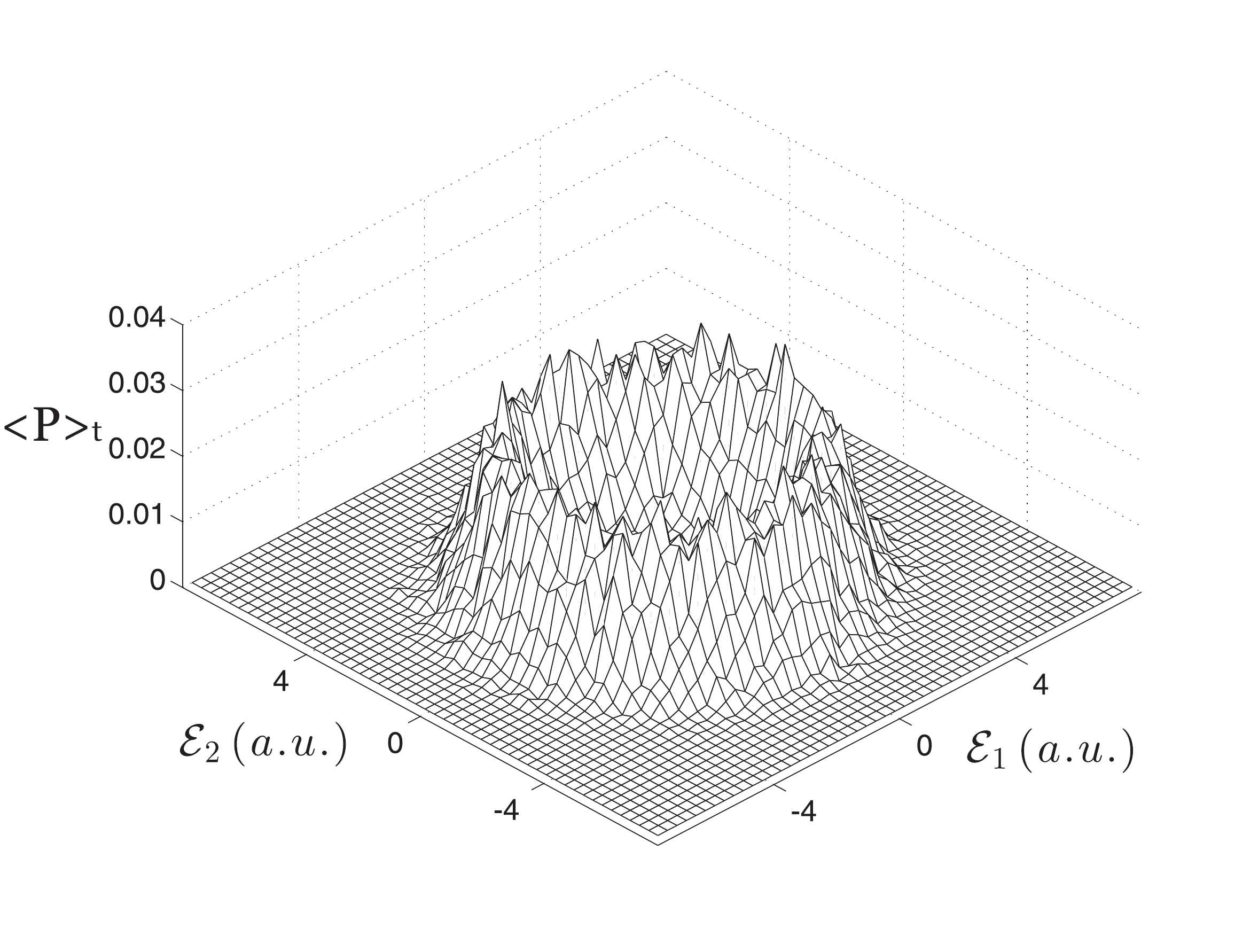}}
\vspace {-0.3cm}\,    
\hspace{-2cm}\caption{\label{F6}\small{Simulation of the measurement of a circularly polarized coherent state (upper figure) and reconstructed probability distribution (lower figure).    }}
\end{figure}

\subsection{Application to quantum homodyne detection}

In this subsection we study the application of the electro-optical SU(2) device to measure  the optical field strength probability distribution  of a two-mode quantum state. The knowledge of such a distribution is very useful to both   obtain the so-called  generalized quantum polarization of the state (wich is given by an accumulated probability distribution) and assess its generalized polarization degree \cite{Barral2013}. The measurement of this feature of any $N$-mode quantum state of light is accomplished by means of an homodyne scheme, phase-independent in the case of stationary quantum states, or phase-averaged in the case of time-dependent states. From the theory of QST, it is known that in order to measure a quantum state, a set of linear transformations must be applied to generate a tomographically complete set of observables, a quorum, whose statistical properties are measured \cite{Raymer1999}. In the usual single-mode homodyne detection this is carried out by means of modulation of an local oscillator  (LO) phase which turns out into a rotation in the phase space \cite{Smithey1993}. When two modes are involved, three parameters are necessary to obtain a quorum. Some techniques have been developed to perform this experimentally, as the "Dual-mode-LO" and the generalized rotations in phase space ''GRIPS'' \cite{Raymer1999}, where the set of transformations is applied to a two-mode LO or to the two-mode signal before mixing in a balanced homodyne detector (BHD), respectively. In this paper we apply this last approach to the detection of spatial quantum states of light in the optical field-strength space $\boldsymbol{\mathcal{E}}$. In this case only one free parameter or none will be necessary because of the option of using random measurement, as we will show below. Hence we propose the detection scheme sketched in Figure 4 (solid line): every measurement unit (a stage) is made up by the electro-optic coupler presented in the above Section performing a rotation in the $\mathcal{E}_{1} \mathcal{E}_{2}$-space. Mathematically this is equivalent to perform the transformation (\ref{M2}) with $\Phi=n\pi$ on the input state, leading to:

\begin{equation} \label{CombQuad}
\begin{pmatrix}
\hat{\mathcal{E}}_{3} \\
\hat{\mathcal{E}}_{4} 
  \end{pmatrix}
=
\begin{pmatrix}
     A(\delta) & (-1)^{n}B(\delta)\\
     (-1)^{n+1}B(\delta) & A(\delta)
  \end{pmatrix}
\begin{pmatrix}
\hat{\mathcal{E}}_{1} \\
\hat{\mathcal{E}}_{2} 
  \end{pmatrix}
\end{equation}

\begin{figure*}[t]
\begin{center}
\includegraphics[width=0.85\textwidth]{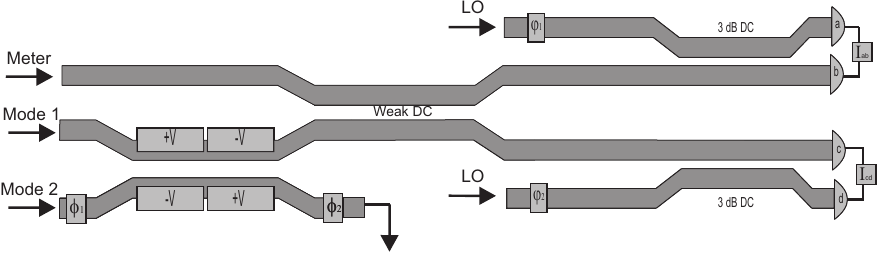}
\vspace {-0.3cm}\,
\hspace{-2cm}\caption{\label{F10}\small{Homodyne detection scheme for weak values measurement of a two-mode quantum state of light. Modes $1$ and $2$ are sent to the Kogelnik-Schmidt coupler where rotations in the field-strength space $\boldsymbol{\mathcal{E}}$ are performed. The upper output mode is weakly coupled in a directional coupler (Weak DC) with a meter mode. The lower ouput mode is sent to a BHD where postselected probabilities are calculated from data and the upper output mode is sent to other BHD where postselected expectation values of the meter conjugate quadrature are calculated.}}
\end{center}
\end {figure*}

This transformation in operator form is given by $\hat{U}_{R}(\chi)=e^{i \chi(\delta)\hat{\sigma}_{y}}$, where $\chi(\delta)= {\rm atan}(B/A)=\Theta/2$ is the effective angle of rotation. The output mode $3$ is sent to the integrated BHD right after, where it is mixed with a strong local oscilator excited in a coherent state $\vert \alpha \rangle$, with $\alpha=\vert \alpha \vert e^{i\psi}$, in the same spatial mode (mode $0$). The output mode $4$ can be used for other purposes or, in the case of a $N$-mode quantum state, it can be the input to the next measurement unit (stage $2$, dashed line, Figure 4)  \cite{Raymer1999}. In the BHD the LO phase $\psi$ performs rotations in the phase space of the output mode $3$, {that is, $\mathcal{E}_{3}\mathcal{P}_{3}$}. Such a rotation is mathematically carried out by the operator $\hat{U}_{LO}(\psi)=e^{-i\psi\hat{n}_{0}}$. From the difference of the $a$ and $b$ photodetectors readings we can obtain statistical information of the state like the moments of the distribution. The mean value of the field and its variance would be for example \cite{Loudon1987},
\begin{eqnarray}\label{Homod}
\langle \hat{\mathcal{I}}_{ab}(\chi, \psi) \rangle \propto 2\vert \alpha \vert \langle \hat{\mathcal{E}}_{3}(\chi, \psi) \rangle, \\
\langle (\Delta\hat{\mathcal{I}}_{ab}(\chi, \psi))^{2} \rangle \propto 4\vert \alpha \vert^{2} \langle (\Delta \hat{\mathcal{E}}_{3}(\chi, \psi) )^{2} \rangle,
\end{eqnarray}
where $\mathcal{\hat{I}}_{ab}$ is the difference of intensities measured by the detectors. However, for a complete characterization of the generalized polarization of the state we have to obtain the total probability distribution $P(\mathcal{E}_{1}, \mathcal{E}_{2})$. This can be accomplished performing sampling to buid up a histogram for every rotation angle $\chi$ and LO phase $\psi$ which gives us an approximate probability distribution of obtaining a value of the field-strength $\mathcal{E}_{3}$:
\begin{equation}\label{P3}
 P(\mathcal{E}_{3}(\chi, \psi))=\langle \mathcal{E}_{1} \mathcal{E}_{2} \vert \hat{U}^{\dag}\, \hat{\rho} \,  \hat{U}\vert \mathcal{E}_{1} \mathcal{E}_{2} \rangle,
 \end{equation}
where $\hat{\rho}$ is the density operator which characterizes the input quantum state, $\hat{\rho}=\vert \Psi \rangle \langle \Psi \vert$ in the case of a pure state, and $\hat{U}=\hat{U}_{LO}(\psi)\,\hat{U}_{R}(\chi)$ the unitary transform performed by the entire detection system. In the case of non-stationary quantum states, we are interested in the accumulated probability distribution over one temporal cycle $\langle P(\mathcal{E}_{1} \mathcal{E}_{2})\rangle_{t}$. This provides us with three methods of sampling this built-up quantum probability distribution: the first one is to control $\psi$ and to vary $\chi$ without following any order. We can call this standard homodyne detection since we set discretely the time (phase) and measure in any field-strength $\mathcal{E}_{3}$, covering all the $\mathcal{E}_{1} \mathcal{E}_{2}$-space and a temporal cycle; the second one is based on setting $\chi$ and uniformly randomize $\psi$, which it could be called phase-random homodyne detection, given that we measure at random times (phases) on the discretely-varied field-strength $\mathcal{E}_{3}(\chi)$, giving us a time-weighted average probability:
\begin{equation}\label{time}
\langle P(\mathcal{E}_{3}(\chi) \rangle_{t}=\frac{1}{2 \pi} \int_{0}^{2\pi} P(\mathcal{E}_{3}(\chi,\psi))\, d\psi
\end{equation}
The third one would be to randomize both $\psi$ and $\chi$, leading to high simplification of the measurement procedure. The discrete nature of the variation of the two parameters and the same sampling process make this procedure inherentely approximated, such that enough resolution has to be reached to obtain a satisfactory outcome. Next, for the sake of clarity, we show in Figure 5 (upper Figure) an standard homodyne detection simulation of the quantum polarization of a circularly polarized coherent state given by $\vert L \rangle = \vert \alpha\rangle_{1} \vert i\alpha \rangle_{2}$, with $\vert \alpha \vert=4$, where $\chi$ has been randomized. This simulation has been carried out creating $10^5$ random points by a Monte Carlo method \cite{Martinez2002}. The outcome obtained performing the experimental procedure explained above would be similar to this. In Figure 5 (lower Figure) we show the probability distribution (\ref{time}) reconstructed from the data sampled in Figure 5 (upper Figure). Fitting this surface allows us to recover the parameters defining the state, as the photon number in each mode and the quantum noise. In this case we obtain $\langle\hat{\mathcal{E}}_{1}\rangle=4.001$, $\langle\hat{\mathcal{E}}_{2}\rangle=4.003$, $\langle\Delta \hat{\mathcal{E}}_{1}\rangle=1.035$ and $\langle\Delta \hat{\mathcal{E}}_{2}\rangle=1.033$, which agree to a great extent with the parameters defining the quantum state $\vert L \rangle$. These data could be be used as well for working out the generalized polarization degree of the state, as shown in \cite{Barral2013}. Of course, the procedure above presented can be also carried out with non-gaussian states, as we show in the following section.

It is important to outline that this scheme is able to be generalized to $N$-mode input states, as the unused output of the directional coupler can be mixed with a third mode and measured by another BHD as well, and so on and so forth up to the $N$-mode. {In Figure 4 is sketched the scheme for the case 3-mode}. So we would need one local oscillator mode and one BHD per input mode to be measured. Moreover, this scheme could be used to obtain the Wigner function on chip by using the transformation (\ref{M2}), repeated measurements of $P(\mathcal{E}_{3}(\chi, \psi, \phi))$ for different combinations of the parameters $(\chi, \psi, \phi)$ and reconstruction algorithms as inverse linear transform techniques or statistical inference \cite{Lvovsky2008}. Additionally, using the phase-averaged distributions (\ref{time}), we can reconstruct the photon number statistics of any input $N$-mode state following \cite{Munroe1995}. In the next section we show a faster way of obtaining complete information about the quantum state, measuring at the same time amplitude and phase of the input wavefunction.



\subsection{Application to  quantum detection of weak values}

Another possible application of this device is  the measurement of the wavefunction by means of weak values. Quantum state reconstruction using this approach has become a topic of great interest since its introduction in \cite{Aharonov1988}. This formalism is founded on the observed outcomes of weak measurements, given by:
\begin{equation}\label{Weak}
A_{w}=\frac{\langle \Psi_{j}\vert \hat{A}\,\vert \Psi\rangle }{\langle \Psi_{j}\vert \Psi\rangle},
\end{equation}
where $A_{w}$ is the complex weak value of the observable given by the quantum operator $\hat{A}$ for a quantum state $\vert \Psi\rangle$ postselected in $\vert \Psi_{j}\rangle$  \cite{Jozsa2007}. This value is dependent on two facts: imposing postselection into a given final state and a weak interaction between the mesurement apparatus (the meter) and the quantum system of interest. For a complete and detailed review see \cite{Aharonov2005}.

Recently, a weak values scheme was proposed to reconstruct the quantum state of an optical single-mode field \cite{Fischbach2012}. This implementation is based on a weak beam splitter interaction between two single-mode quantum states, the one to be measured $\vert \Psi\rangle$ and the other a gaussian state acting as a meter {$\vert \mu\rangle$}. Following this approach the quantum state is reconstructed in the optical field momentum $\mathcal{P}$ basis, where the wavefunction is given by:
\begin{equation}\label{WFP}
\Psi(\mathcal{P})=\vert \Psi (\mathcal{P})\vert\,e^{i\,\varphi(\mathcal{P})}.
\end{equation}
This scheme is also suited to be applied in integrated optics. In this case homodyne detection is carried out on each output guide of a directional coupler with weak coupling parameter $\Gamma_{w}=\kappa L$, where $L$ is the coupling length (or $\chi_{w}(\delta)$ if our electro-optic coupler is used for this task), taking postselection probabilities in one guide $P(\mathcal{P})=\vert \Psi (\mathcal{P}) \vert^{2}$ (strong measurement) and postselected expectation values in the conjugate quadrature of the meter $E^{(\mathcal{P})}[\hat{\mathcal{E}}_{\mu}]=-\Gamma_{w}\,\partial\varphi(\mathcal{P})/\partial\mathcal{P}$ (weak measurement) in the other guide. From the strong measurement we obtain the amplitude of (\ref{WFP}), and from the weak measurement its phase by means of:
\begin{equation}\label{PWP}
\varphi(\mathcal{P})=-\frac{1}{\Gamma_{w}}\,\int^{\mathcal{P}} E^{(\mathcal{P'})}[\hat{\mathcal{E}}_{\mu}] \,dP'.
\end{equation}

This section is devoted to extend this weak detection scheme to two-mode (or $N$-mode) spatial quantum states of light. So, on one hand, since we are interested in measuring spatial quantum states, we have translated this scheme to photonics by the use of a fixed directional coupler, or our reconfigurable device allowing us sharper selection of the strength of the interaction, as weak interaction system. On the other hand, as our aim is the study of two-mode quantum states, performing rotations in the $\mathcal{P}_{1}\,\mathcal{P}_{2}$ plane by means of our device, which, mathematically, consists of  substituting $\mathcal{E}$ by $\mathcal{P}$ in equation (\ref{CombQuad}), we can obtain full amplitude and phase information for every rotated angle $\chi$. In Figure 6 we show a sketch of the circuit proposed for this weak measurement-based detector. The principal advantage of this scheme is the quicker acquisition and simpler analysis of data with respect to QST. But unfortunately this approach presents some drawbacks. The main drawback, inherent to the method, is its inability to reconstruct the phase for those values of $\mathcal{P}$ with low probability, as it can be seen in Figures 7 and 8: the lower the probability, the poorer the reconstructed phase. This can be overlooked for the far values of the field, but it is an unavoidable problem when the wavefunction presents gaps \cite{Fischbach2012}. A second drawback, dependent in this case on our design, is the inability of reconstruct quantum states with only angular dependence, as the postselected expectation value will be zero for every rotation angle $\chi$. In these cases QST has to be chosen. It is important to outline that a hybrid scheme of QST and weak measurement is possible, leaving QST for those states or values of $\chi$ not suitable to be measured by the weak scheme. 

Figures 7 and 8 show theoretical and simulated reconstructions of the weak measurements carried out over a NOON-type state $\vert L \rangle=\vert 2\,0\rangle + i \vert 0\,2\rangle$ by means of phase-random homodyne detection. We have used a weak coupling parameter $\Gamma_{w}=0.05$ and $N=10^5$ data points. Figure 7 shows the simulated and theoretical probability for two rotation angles $\chi=0, \pi/3$ (upper Figure) and the total data sampled in the $\mathcal{P}_{1}\,\mathcal{P}_{2}$-space (lower Figure). Likewise, the phase is reconstructed integrating the expectation value of the meter $E^{(\mathcal{P}_{3})}[\mathcal{E}_{\mu}]$, in our case the vaccuum $\vert \mu = 0 \rangle$, over $\mathcal{P}_{3}$. As it can be seen in the upper Figure 8, where theoretical and reconstructed phases for two rotation angles $\chi=0, \pi/3$ are shown, they highly agree for values of $\vert\mathcal{P}_{3}\vert<2.5$ where the probability of the quantum state is high (upper Figure 7), as it was discussed above. In the lower Figure 8 we show the reconstructed joint phase of the input quantum state obtained from individual measurements as those depicted on upper Figure 8. Note that as only relative phases are physically meaningful, we have chosen the phase origin at $(\mathcal{P}_{1},\,\mathcal{P}_{2})=(0,0)$ and it acts as the lower bound of integral (\ref{PWP}). As it can be seen, the phase of  $\vert L \rangle$ varies smoothly unless for $\chi=\pm \pi/4$, where phase jumps appear and the method does not work. To solve this, we have interpolated the data at these planes. Likewise, in lower Figure 8 we have dismissed the reconstructed phases for values of $\mathcal{P}_{1},\,\mathcal{P}_{2} >3$, as those values are far from the theoretical. 

Hence, this procedure gives us full characterization of the joint quantum wavefunction. It is important to say that the reconstructed wavefunction obtained is the same as that in the field-strength space $\boldsymbol{\mathcal{E}}$ because of the temporal invariance of the quantum state, but in general it can be obtained by a Fourier Transform. In the case of a non-stationary quantum state, standard homodyne detection would be required for reconstructing the amplitude and phase of the state in every point of a temporal cycle.

\begin{figure}[h]
  \centering
    \subfigure{\includegraphics[width=0.47\textwidth]{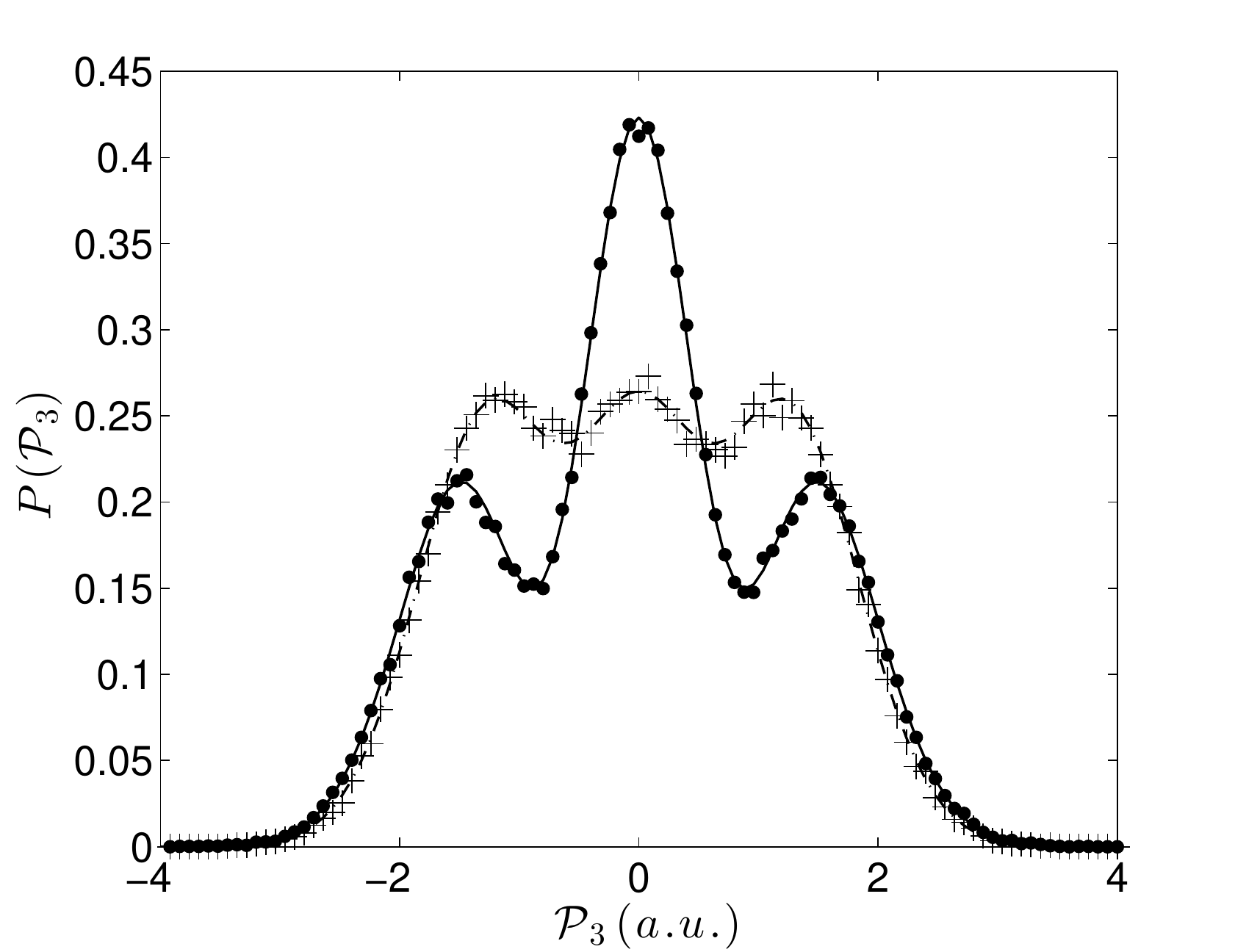}}
    \subfigure{\includegraphics[width=0.47\textwidth]{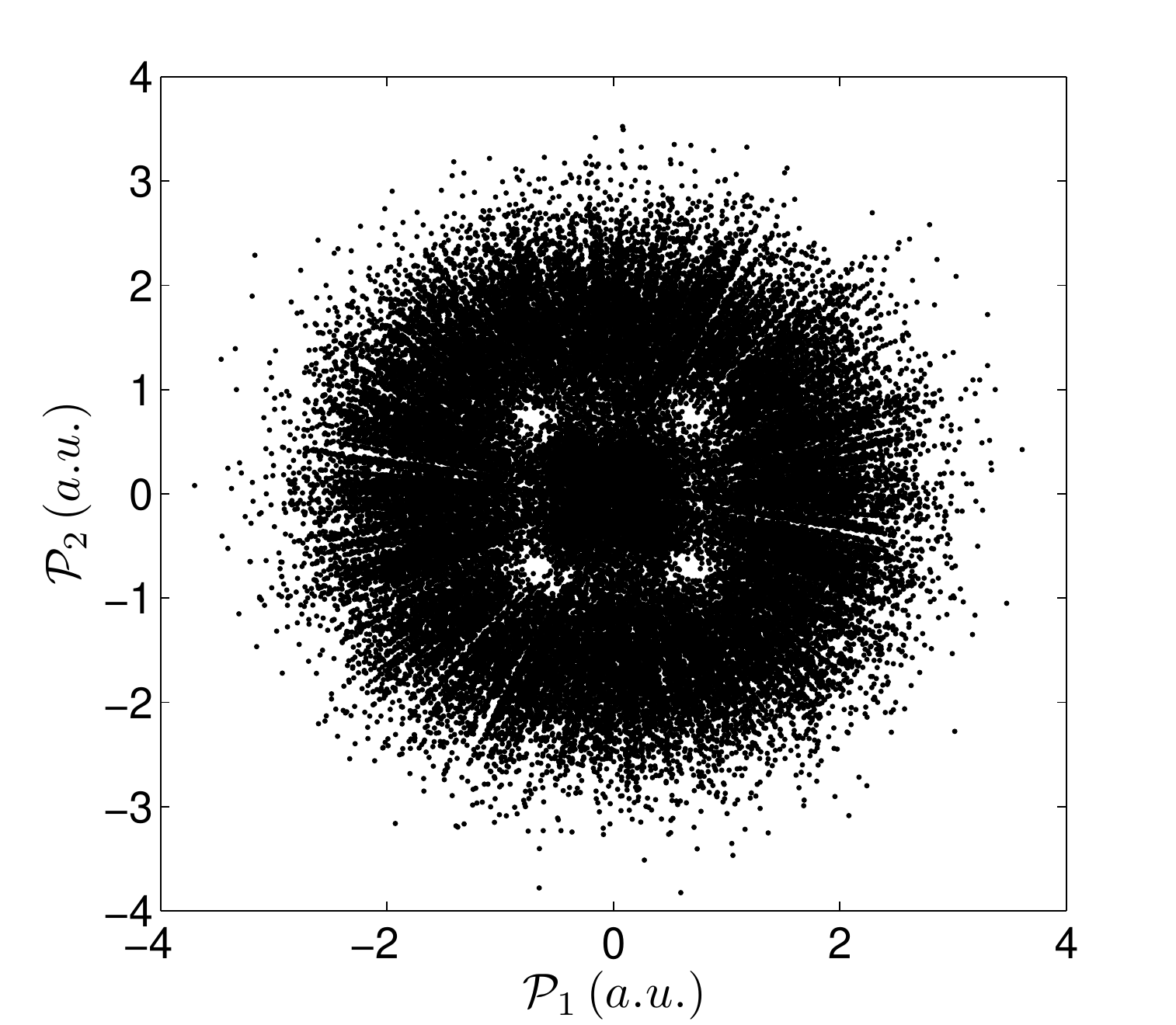}}
\vspace {-0.3cm}\,
\hspace{-2cm}\caption{\label{F8}\small{Probability corresponding to a NOON-type state with $N=2$. Upper Figure: theoretical and simulated probability corresponding to two values of $\chi$: theoretical (solid line) and simulated (dots) probability for $\chi=0$ and theoretical (dash-dot line) and simulated (crosses) probability for $\chi=\pi/3$. Lower Figure: total data sampled for probability reconstruction. }}
\end{figure}

\begin{figure}[h]
  \centering
    \subfigure{\includegraphics[width=0.47\textwidth]{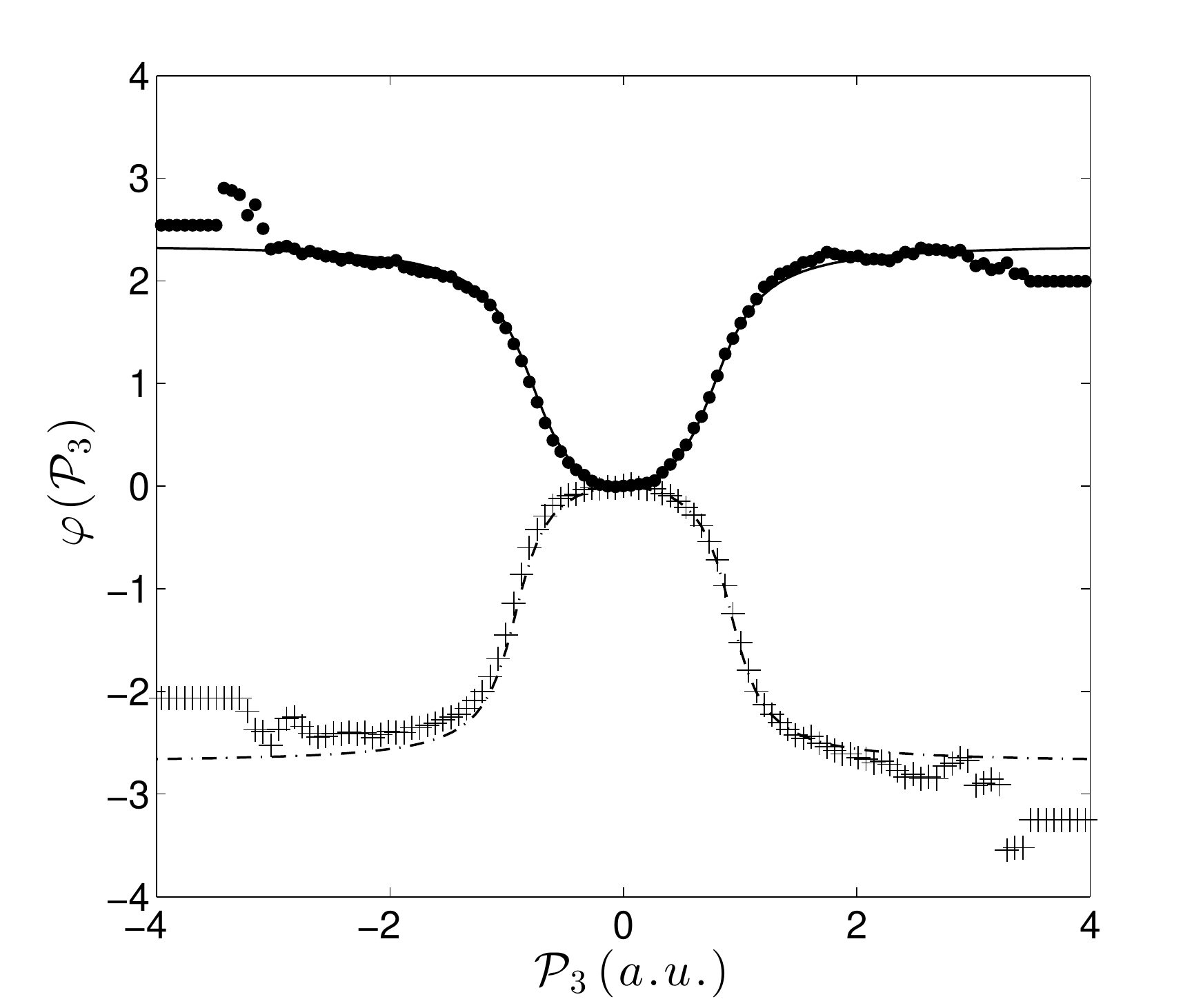}}
    \subfigure{\includegraphics[width=0.5\textwidth]{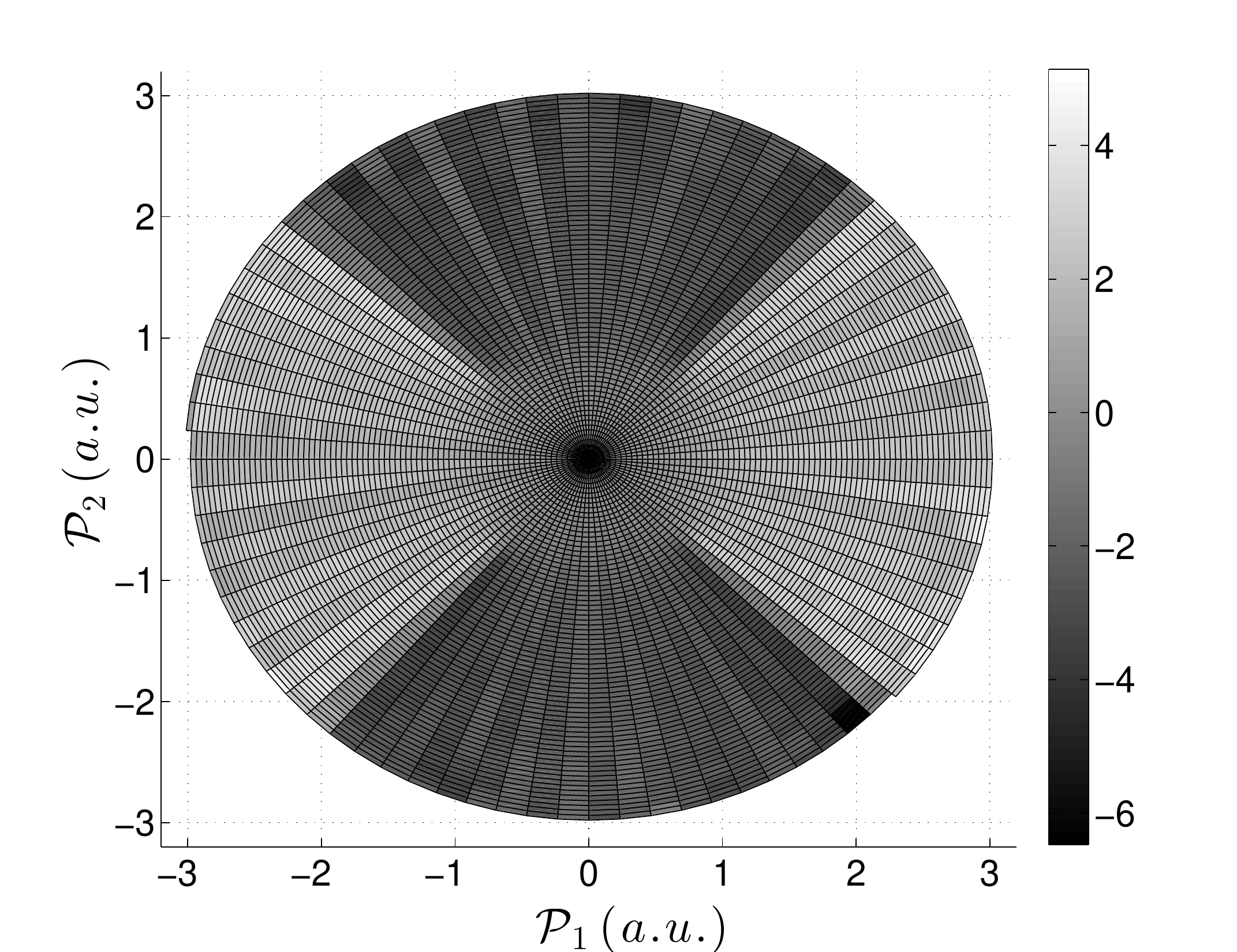}}
\vspace {-0.3cm}\,
\hspace{-2cm}\caption{\label{F9}\small{ Phase corresponding to a NOON-type state with $N=2$. Upper Figure: theoretical and simulated phase corresponding to two values of $\chi$: theoretical (solid line) and simulated (dots) phase for $\chi=0$ and theoretical (dash-dot line) and simulated (crosses) phase for $\chi=\pi/3$.  Lower Figure: reconstructed joint phase.}}
\end{figure}

\section{Summary}
In this work we have studied the detection of two-mode spatial quantum states of light with an homodyne on-chip scheme. We have designed a device capable of carry out reconfigurable SU(2) and SO(2,R) transformations on spatial modes by means of a directional coupler built in an electro-optical material, extensible to N-mode input quantum states by nesting, and we have compared its performance with other current schemes. Finally we applied it to the measure of generalized quantum polarization and reconstruction of the wavefunction using weak values.


\begin{thebibliography}{50}
\bibitem{OBrien2009} O'Brien, J.L., Furusawa A. and Vukovic, J. 2009 {\it Nature Photon.} {\bf 3} 687
\bibitem{Politi2008} Politi, A. et al. 2008 {\it Science} {\bf 320} 646
\bibitem{Mathews2009} Mathews J.C.F. et al. 2009 {\it Nature Photon.} {\bf 3} 346
\bibitem{Smith2009} Smith B.J. et al. 2009 {\it Opt. Express} {\bf 17} (16) 13516
\bibitem{Shadbolt2012} Shadbolt, P.J. et al. 2012 {\it Nature Photon.} {\bf 6} 45
\bibitem{Bonneau2012a} Bonneau, D. et al. 2012 {\it New J. Phys.} {\bf 14} 045003
\bibitem{Metcalf2014} Metcalf, B.J. et al. 2014 {\it Nature Photon.} {\bf 8} 770
\bibitem{Silverstone2014} Silverstone, J. et al. 2014 {\it Nature Photon.} {\bf 8} 104
\bibitem{Humphreys2014} Humphreys, P.C. et al 2014 {\it Opt. Express} {\bf 22} (18) 21719
\bibitem{Martin2012} Martin A. et al. 2012 {\it New J. Phys.} {\bf 14} 025002
\bibitem{Jin2014} Jin H. et al. 2014 {\it Phys. Rev. Lett.} {\bf 113} 103601
\bibitem{Bonneau2012} Bonneau, D. et al. 2012 {\it Phys. Rev. Lett.} {\bf 108} 053601
\bibitem{Saglamyurek2011} Saglamyurek, E. et al. 2011 {\it Nature} {\bf 469} 512
\bibitem{Santori2002} Santori, C. et al 2002 {\it Nature} {\bf 419} 594
\bibitem{Wang2014} Wang, J. et al. 2014 {\it Optics Comm.} {\bf 327} 49
\bibitem{Sahin2015} Sahin, D. et al 2015 {\it IEEE J. Sel. Top. Quantum Electron.} {\bf 21} (2) 3800210
\bibitem{Nielsen2010} Nielsen, M.A. and Chuang, I.L. 2010 Quantum Computation and Quantum Information, Cambridge University Press
\bibitem{Lvovsky2008} Lvovsky, A.I. and Raymer, M.G. 2008 {\it Rev. Mod. Phys.} {\bf 81} 299
\bibitem{Smithey1993} Smithey, D.T. et al. 1993 {\it Phys. Rev. Lett.} {\bf 70} (9) 1244
\bibitem{Opatrny1997} Opatrny, T., Welsch D.-G. and Vogel, W. 1997 {\it Optics Comm.} {\bf 134} 112
\bibitem{Raymer1996} Raymer, M.G., McAlister, D.F. and Leonhardt, U. 1996 {\it Phys. Rev. A} {\bf 54} (3) 2397
\bibitem{Aharonov1988} Aharonov, Y., Albert D. Z. and Vaidman, L. 1988 {\it Phys. Rev. Lett.}{\bf 60} 1351
\bibitem{Lundeen2011} Lundeen J. S. et al. 2011 {\it Nature} {\bf 474} 188
\bibitem{Linares2011} Li$\tilde{\rm{n}}$ares, J. et al. 2011 {\it J. Mod. Opt. } {\bf 58} 711
\bibitem{Barral2013} Barral D., Li$\tilde{\rm{n}}$ares J. and Nistal M. C. 2013 {\it J. Mod. Opt. } {\bf 60} (12) 941
\bibitem{Luis2013} Luis, A. and Sanz, A.S. 2013 {\it Phys. Rev. A} {\bf 87} 063844
\bibitem{Braunstein2005} Braunstein, S.L. and van Loock, P. 2005 {\it Rev. Mod. Physics} {\bf 77} 513
\bibitem{Kogelnik1976} Kogelnik, H. and Schmidt, R.V. 1976 {\it IEEE J. Quantum Electron.} {\bf 12} (7) 396
\bibitem{Luks2002} Luks, A. and Perinova, V. 2002 {\it Progress in Optics} {\bf 43} 295
\bibitem{Linares2008} Li$\tilde{\rm{n}}$ares J., Nistal M. C. and Barral D. 2008 {\it New J. Phys. }{\bf 10} 063023
\bibitem{Linares2012} Li$\tilde{\rm{n}}$ares J., Barral D. and Nistal M. C. 2012 {\it J. Nonlin. Opt. Phys. Mat. }{\bf 21} 1250032
\bibitem{Pfeifle2012} Pfeifle, J. et al. 2012 {\it Opt. Express} {\bf 20} (14) 15359
\bibitem{Wooten2000} Wooten, E.L. et al. 2000 {\it IEEE J. Sel. Top. Quantum Electron.} {\bf 6} (1) 69
\bibitem{Walker2012} Walker, R.G. and Heaton, J., 2012 {\it Gallium arsenide modulator technology} in Broadband Optical Modulators-Science, Technology and Applications, Eds. A. Chen and E. J. Murphy, Boca Raton, CRC Press
\bibitem{Reck1994} Reck, M. et al. 1994 {\it Phys. Rev. Lett.} {\bf 73} (1) 58
\bibitem{Yariv1975} Yariv, A., 1975 Quantum Electronics, New York, John Wiley \& Sons
\bibitem{Kanno2010} Kanno, A. et al. 2010 {\it IEICE Electron. Express} {\bf 7} 817
\bibitem{Raymer1999} Raymer, M.G. and Funk, A.C. 1999 {\it Phys. Rev. A} {\bf 61} 015801
\bibitem{Loudon1987} Loudon, R. and Knight, P.L. 1987 {\it J. Mod. Opt.} {\bf 34} (6-7) 709
\bibitem{Martinez2002} Martinez W. and Martinez A. 2002 Computational Statistics Handbook with Matlab, Chapman \& Hall/CRC
\bibitem{Munroe1995} Munroe M. et al. 1995 {\it Phys. Rev. A} {\bf 52} (2) R924
\bibitem{Jozsa2007} Jozsa, R. 2007 {\it Phys. Rev. A} {\bf 76} 044103
\bibitem{Aharonov2005} Aharonov, Y. and Rohrlich, D. 2005 Quantum Paradoxes, Weinheim, Wiley-VCH
\bibitem{Fischbach2012} Fischbach, J. and Freyberger, M. 2012 {\it Phys. Rev. A} {\bf 86} 052110
\end{thebibliography}
\end{document}